\documentclass[11pt, fleqn]{article} 

\usepackage{amsmath}
\usepackage{amsfonts}
\usepackage{graphicx}
\usepackage[section]{placeins}
\usepackage{multirow}
\usepackage{dcolumn}

\newcommand{\eigdecomp}{eigen-decomposition}

\newcommand{\dt}{\delta t}

\newcommand{\tr}{\;\text{tr}}
\newcommand{\id}{\mathbb{I}}
\newcommand{\iMax}{i_\text{max}}
\newcommand{\bfr}{\mathbf{r}}
\newcommand{\sfP}{\mathsf{P}}
\newcommand{\SigmaEff}{\Sigma_{\text{eff}}}
\newcommand{\ev}{e}  % The eigenvalues
\newcommand{\avg}[1]{\left< #1 \right>}

\newcommand{\figpath}{./figures}
\begin{document}
\vspace*{5ex}
\begin{center}
{\LARGE\bf The empirical properties of large covariance matrices}\\[3ex]
\vspace*{8ex}
\parbox{0.4\textwidth}{\renewcommand{\baselinestretch}{1.0}\normalsize
{\bf Gilles Zumbach} \\[2ex]
RiskMetrics Group \\
Av. des Morgines 12 \\
1213 Petit-Lancy\\
Geneva, Switzerland\\[1ex]
gilles.zumbach@riskmetrics.com}

\vspace{5ex}
keywords: Covariance matrix, spectrum, spectral density 

\vspace*{8ex}
\today
%November 4, 2008

\vspace{8ex}

\end{center}
\begin{abstract}
The salient properties of large empirical covariance and correlation matrices are studied for three datasets of size 54, 55 and 330.
The covariance is defined as a simple cross product of the returns, with weights that decay logarithmically slowly.
The key general properties of the covariance matrices are the following.
The spectrum of the covariance is very static, except for the top three to ten eigenvalues, and decay exponentially  fast toward zero.
The mean spectrum and spectral density show no particular feature that would separate ``meaningful'' from ``noisy'' eigenvalues.
The spectrum of the correlation is more static, with three to five eigenvalues that have distinct dynamics.
The mean projector of rank $k$ on the leading subspace shows instead that most of the dynamics occur in the eigenvectors, including deep in the spectrum.
Together, this implies that the reduction of the covariance to a few leading eigenmodes misses most of the dynamics,
and that a covariance estimator correctly evaluates both volatilities and correlations.\end{abstract}

\setcounter{footnote}{1}
\renewcommand{\thefootnote}{\fnsymbol{footnote}}
\footnotetext{The author wishes to thank Romain Cosandey with whom part of this work was done, and Chris Finger for many discussions and carefull review of the manuscript.}
\renewcommand{\thefootnote}{\arabic{footnote}}
\setcounter{footnote}{0}

\newpage
\section{Introduction}
%------------------------------------------
In a given investment universe, the covariance matrix encapsulates a wealth of information.
Part is related to the volatility of each component, and part to the mutual dependencies of the components as measured by the correlations.
Aiming at practical applications, the number of time series $N$ should be large, up to one thousand or more.
The corresponding covariance is of size $N\times N$, resulting in a huge amount of information.
Moreover, the dynamics of the covariance are important, as the volatilities and correlations evolve with time.
Therefore, we are faced with $N^2$ time series.
This is clearly difficult to grasp, and the raw covariance matrix should be transformed into more palatable quantities.

Because the covariance matrix is symmetric, a natural tool is an \eigdecomp.
With the time dependency, the eigenvectors and eigenvalues are themselves time series which contain the same information as the raw covariance.
For large $N$, the eigenvalues have some generic properties that allow us to better understand, and possibly to summarize, the information contained in the covariance (or correlation) matrices.
This is the main goal of this paper.

The covariance matrix appears in many computations in finance.
Common examples are in Monte Carlo schemes for instrument pricing or for risk evaluation, in optimal portfolio allocation or in process inference using a log-likelihood maximization.
For the last two examples, the {\it inverse} of the covariance is needed (the inverse square root for process inference).
As the number of time series $N$ becomes large, the covariance contains null eigenvalues.
In this case, the inverse is not defined, and a proper regularization scheme should be used.
This occurs always when $N$ is larger than the memory length $\iMax$ used to compute the covariance.
For many practical applications, the memory length is of the order of one to two years ($\iMax = 260$ to $\iMax = 520$), whereas the number of time series can be of the order of thousands.
As shown in this paper, similar practical problems occur even when the covariance should be (mathematically) regular.
This happens because the spectrum decays exponentially fast toward zero, leading to very large values in the inverse.
Essentially, for $N$ large, the presence of many small (or null) eigenvalues makes the inverse covariance badly defined, and a regularization scheme should be used even when the covariance is in principle regular.

It is natural to define the regularization scheme using the \eigdecomp.
An important issue is then the number of eigenvalues to include in the inverse covariance.
In order to select a cutoff, the ``meaningful'' eigenvalues should be separated from the ``noise-induced'' eigenvalues.
This is essentially the line followed by Principal Components Analysis (PCA) and by factor models.
Another idea is to use random matrix theory to find the edge of the noise induced spectrum.
As we will see, no simple threshold can be found: % by investigating the spectrum, the spectral density and the subspaces span by the eigenvectors.
the covariance structure is more of a gradual crossover from meaningful to noise-induced.

When estimating covariance matrices, the question of the best covariance estimator arises, as well as of the best forecast.
Several paths can be followed to define the covariance matrix.
First, the covariance can be computed as the natural extension of the univariate case, essentially replacing $r^2$ in the univariate formulas by the product $r_\alpha\cdot r_\beta$ in the multivariate case.
This is the approach taken in the present paper, using slow decaying weights that produce the best univariate forecast.
This route imposes the least structure on the covariance.

Second, the volatility and correlation can be separated, and each part evaluated with a specific and optimal formula.
The underlying idea is that the volatilities have fairly fast dynamics, while the correlations are more stable.
Correspondingly, the optimal estimators for both quantities should be different.
This idea has been pursued in the GARCH context by \cite{Bollerslev.1990} and subsequent papers which use a constant correlation matrix, or a correlation with slow dynamics.
This issue can be formulated as the identification of two different characteristic time scales in the covariance: a short one for the fast evolution of the volatilities (from a few days to a few months), and a long one corresponding to the correlation (at least one year).
This paper investigates the characteristic time of the fluctuations using lagged correlations, for the covariance and correlation matrices, and for the projectors on the leading subspaces.
%By pushing the underlying argument for the separation of volatilities and correlations, we can take the correlations to be constant,  bringing to the limit the idea of very stable correlations.
%The CCC-GARCH process of \cite{Bollerslev.1990} follows this approach for multivariate GARCH modeling.

Third, more structure can be imposed on the covariance, for example by a factor model, or by a Bayesian method with an given prior for the covariance matrix \cite{Ledoit-Wolf.2003}.
A comparison of various structures for the covariance matrix is presented in \cite{BrinerConnor.2008} for a universe of UK equities, with the aim at volatility forecasting.
This route implies that some extra information exists about the covariance matrix, and this information is dominant and stable through time.
The present paper investigates the stability of the structures present in the covariance matrix by analyzing the dynamics of the eigenspaces of increasing ranks.

This paper focuses on the empirical properties of the covariance and correlation for large $N$, using the spectrum, the spectral density and the subspaces span by the eigenvectors.
The study is done for three datasets of size 54, 55 and 330.
This allows to extract generic features of the covariance.
A subsequent paper \cite{Zumbach.inferenceOnProcesses} uses these properties to study various definitions for the regularization of the inverse, and the implications for process inference.

The structure of this paper is the following.
The next section introduces the relevant definition and theoretical material.
The datasets are described in Section~\ref{sec:datasets}.
Sections~\ref{sec:covarianceSpectrum} and \ref{sec:correlationSpectrum} analyze the dynamic of the spectrum
for the correlation and covariance matrices.
The spectral density of the correlation is investigated in Section~\ref{sec:spectralDensity_correlation}, and relations are made with the random matrix theory.
Section~\ref{sec:spectrum_covariance} analyzes the mean spectrum and spectral density of the covariance.
In Section \ref{sec:meanProjector}, the structure of the eigenvectors is investigated through the mean projectors on the leading subspaces with given ranks, while Section~\ref{sec:projectorDynamics} focuses on the dynamics of the projectors.

\section{Theoretical framework}
%-------------------------------------
\label{sec:generalFramework}

For a time series $x(t)$ with a daily time increment $\dt$, the daily return is defined as
\begin{equation}
    r(t) = x(t) - x(t-\dt).  \label{eq:baseModel_r}
\end{equation}
The mapped price $x$ is $x(t) = \ln(p(t))$ for stock, stock indexes, FX and commodities.
For interest rates, $x$ corresponds to the rate at a fixed time to maturity\footnote{More precisely, for the interest rate $R$, the mapped price is $x = \log(1 + R/R_0)$ with $R_0$ = 4\%. This mapping decouples the volatility from $R$; see \cite{Zumbach.RM2006_fullReport}. This mapping introduces a small correction on the returns for interest rates.}.

We consider the class of covariance matrices $\SigmaEff$ that are the cross product of the past return vectors
\begin{equation}
   \SigmaEff(t) = \sum_{i = 0 }^{\iMax} \lambda (i)\,\bfr(t-i\,\dt) \; \bfr'(t-i\,\dt)   \label{eq:sigmaEffCrs}
\end{equation}
with $\bfr$ a column vector and $\bfr'$ its transpose.
The weight for the past returns $\lambda(i)$ obeys the sum rule $\sum_i \lambda(i) = 1$.
Common choices for the weights are equal weights (i.e. a rectangular window with equal weights), exponential weights, and long-memory weights \cite{Zumbach.RM2006_fullReport}.
The exponential weights decay as $\lambda(i) \simeq \mu^i = \exp(i\:\dt/\tau)$, and are equivalent to an exponential moving average of the past returns.
The long-memory weights decay logarithmically slowly, with $\lambda(i) \simeq 1 - \log(i\:\dt)/\log(\tau_0)$,
and they correspond to the volatility structure presents in the financial time series.
A corresponding process with long-memory weights reproduces the long memory observed in the empirical lagged correlation of the (univariate) volatility \cite{Zumbach.LongMemory}.
Consequently, a volatility forecast based on this process with long-memory weights delivers consistently better forecasts than the other typical choices (equal weights or exponential).
An extensive empirical analysis in \cite{Zumbach.RM2006_fullReport} shows that the same parameter values can be used for {\it all financial time series}, with a decay parameter $\tau_0$ of the order of six years.
The recursion equations used to define the $\lambda(i)$ and the parameter values are given in \cite{Zumbach.RM2006_fullReport}.
Because the long-memory weights better describe the data, we select them for this work.
Yet the detailed shape of the kernel has a minor impact on the salient results.

For symmetric matrices, the \eigdecomp~is
\begin{equation} \label{eq:regularization_projected}
  \SigmaEff = \sum_{\alpha = 1}^{N} \ev_\alpha \;\mathbf{v}_\alpha \mathbf{v}_\alpha'
\end{equation}
where the eigenvalues $\ev_\alpha = \ev_\alpha(t)$ and eigenvectors $\mathbf{v}_\alpha = \mathbf{v}_\alpha(t)$ are time dependent, and the eigenvectors $\mathbf{v}_\alpha$  are orthogonal.
For convenience, the eigenvalues are ordered by decreasing values such that $\ev_\beta \leq \ev_\alpha$ for $\beta > \alpha$.

The logarithmic mean eigenvalues of the covariance and correlation matrices are defined by
\begin{equation}
 \log( \avg{\ev_\alpha}) = \frac{1}{T}\sum_{t=1}^T \log(\ev_\alpha(t)).
\end{equation}
The spectrum of the covariance matrix is investigated in Section~\ref{sec:spectrum_covariance}.
The spectral density in the interval $[\lambda - \delta\lambda, \lambda + \delta\lambda]$ is computed by
 \begin{equation}
    \rho(\lambda) = \frac{1}{2\;\delta\lambda} \;\frac{1}{N\:T}\;
	\sum_{t=1}^T \sum_\alpha \chi(\lambda - \delta\lambda < \ev_\alpha < \lambda + \delta\lambda).
\end{equation}
where $\chi(x)$ is the characteristic function of $x$ (1 if $x$ is true, 0 if $x$ is false).
The spectral density $\rho(\lambda)$ measures the time average density of the eigenvalues around $\lambda$.
The normalization is defined such that $\int \rho(\lambda)\; d\lambda = 1$.
In the theoretical computation related to random matrices, the time average is replaced by an ensemble average.

The spectrum and eigenvectors of the covariance matrix need to be studied in order to understand the properties of such large matrices.
The canonical picture in finance is that a few leading eigenvalues dominate, corresponding to broadly defined directions, like ``the market'', ``value versus growth'', or the overall ``interest rate level''.
In mathematical terms, the canonical picture is that the leading eigenvalues $\ev_\alpha$ correspond to stable directions for the corresponding vectors $v_\alpha$.
In order to measure this stability, the mean eigenvectors are not interesting as they converge to zero.
The cause is a sign indeterminacy for the eigenvector: $v_\alpha$ and $-v_\alpha$ have the same eigenvalue.
A second problem is the crossing of the eigenvalues, leading to an apparent abrupt change of the eigenvector with a given rank.
The object that is better behaved for our purpose is the projector on the leading subspace.
The projectors $\sfP_k(t)$ on the leading subspace of rank $k$ is
\begin{equation}
\label{eq:parallelProj}
    \sfP_k = \sum_{\alpha=1}^{k} \mathbf{v}_\alpha \mathbf{v}_\alpha'.
\end{equation}
For a given rank $k$, the mean projector is defined by the time average
\begin{equation} \label{eq:meanProjector}
    \avg{\sfP_k} = \frac{1}{T}\sum_{t=1}^T \sfP_k(t).
\end{equation}
The rank is preserved by the average $\tr\avg{\sfP_k} = k$, but  $\avg{\sfP_k}$ is not a projector as its eigenvalues are between 0 and 1 (and not either 0 or 1).
If the projector is essentially static $\avg{\sfP} \simeq \sfP(t)$, the projector has $k$ eigenvalues 1 and $N-k$ eigenvalues 0.
In the other direction, if the projector dynamics explores fully the available space, the mean projector is proportional to the identity $\avg{\sfP_k} \simeq k/N \;\id$ with all eigenvalues given by $k/N$.

Principal Components Analysis (PCA) is widespread in the investment business.\footnote{See for example \cite{Tsay-05}.}
It uses the spectral decomposition of the covariance matrix to reduce the dimensionality of the original universe to a small number of uncorrelated factors.
A large number of empirical studies (mostly on equities)  on this dimension reduction technique have been done, and tend to show that only a small number of eigenvalues are significant.
%They also often show that beside the usual Fama-French factors, some other new factors are needed, and many such explanatory factors have been proposed.
The picture conveyed by PCA is of a stable leading subspace.
For the mean projector, this picture translates into a set of eigenvalues close to one corresponding to the relevant and stable eigenvectors, with a clear gap separating eigenvalues close to zero corresponding to the idiosyncratic noise.
As the trace is constant, the choice of the rank for the projector alters the eigenvalues.
The relevance of this picture is investigated in Section~\ref{sec:meanProjector}.

\section{The datasets}
%============================================
\label{sec:datasets}
One important motivation for this study is the presence of null eigenvalues in the covariance matrix.
This occurs when the size of the covariance matrix $N$ is larger than the historical depth $\iMax$ in \eqref{eq:sigmaEffCrs}.
In order to be realistic, a history depth of one year (260 business days) has been used.
Moreover, we want to investigate a few investment universes, including one case where the covariance matrix is degenerate, and one where it is not.
These considerations lead to the choice of three datasets.

The {\bf ICM  dataset} (International Capital Market) covers majors
asset classes and world geographical areas,
with a total of 340 time series divided as follows:
 19 commodities %(indexes 1 to 19)
	containing metal %(indexes 1 to 15)
	and gas futures, %(indexes 16 to 19),
 78 foreign exchange rates, %(indexes 20 to 97, all quoted as the number of USD needed to buy one unit of  another currency)
 52 equity indices, %(indexes 98 to 149),
127 interest rates, with maturities at one day, %(indexes 150 to 165)
	one month, %(indexes 166 to 207)
	one year %(indexes 208 to 245)
	and ten years, %(indexes 246 to 276)
and 54 individual stocks from USA, %(indexes 277 to 320)
	 France %(indexes 321 to 326)
	 and Switzerland. % small cap. %(indexes 327 to 330).

The {\bf G10 dataset} covers the largest economies
(European, Japan, and USA).
It contains 55 time series:
 5 commodities,
 6 foreign exchange rates,
 13 equity indices
 and 31 interest rates.% (with maturities at 1 day, 1 month, 1 year and 10 years).

The {\bf USA dataset} focuses on the American economy and is composed of 54 times series:
the S\&P500 and Nasdaq equity indices,
 8 government and swap interest rates %(with maturities at 1 day, 1 month, 1 year and 10 years),
 and 44 individual stocks of the largest US companies on the NYSE.

All time series contain daily prices from 1 January 1999 to 1 January 2008 (nine years), corresponding to a length of 2515 days.
The in-sample investigations are done from 1 January 2000 to 1 January 2008, and one year ($\iMax = 260$ business day) is used to evaluate the effective covariance $\SigmaEff$.
The FX values are a snapshot at a given GMT time, while the values for the other time series correspond to the closing of the market.
Therefore, for the ICM and G10 datasets, a part of the correlations can be induced by the asynchronous nature of the data and the circulation of the information across the globe.
Unfortunately, it is difficult to obtain data with such a large world coverage, with a sufficient time span, and taken at the same world time.
Therefore, although not perfect in term of causality, these datasets are the best that can be used nowadays.
For the USA dataset, all the time series are obtained at the closing of the US market and are therefore synchronous.
The results for the USA set are in line with both world sets, confirming that the asynchronicity is irrelevant for the present paper.

\section{The dynamics of the covariance spectrum}
%-------------------------------------------------------------------------------------------
\label{sec:covarianceSpectrum}

\begin{figure}
  \centering
  \includegraphics[width = 1.0\textwidth]{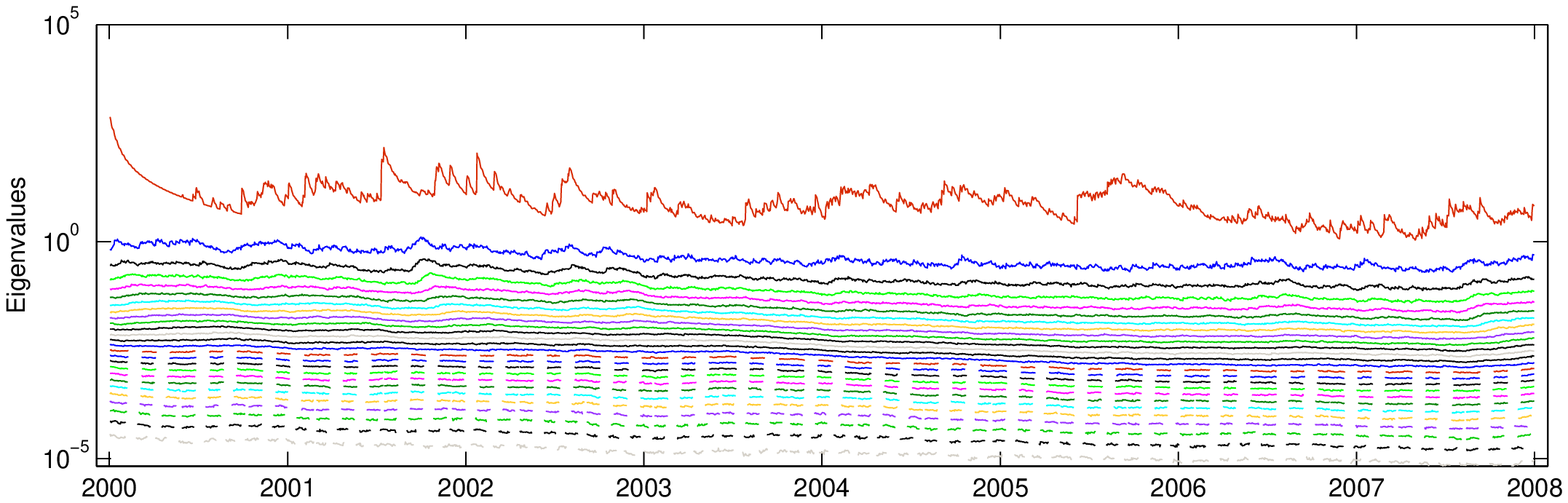}
  \includegraphics[width = 1.0\textwidth]{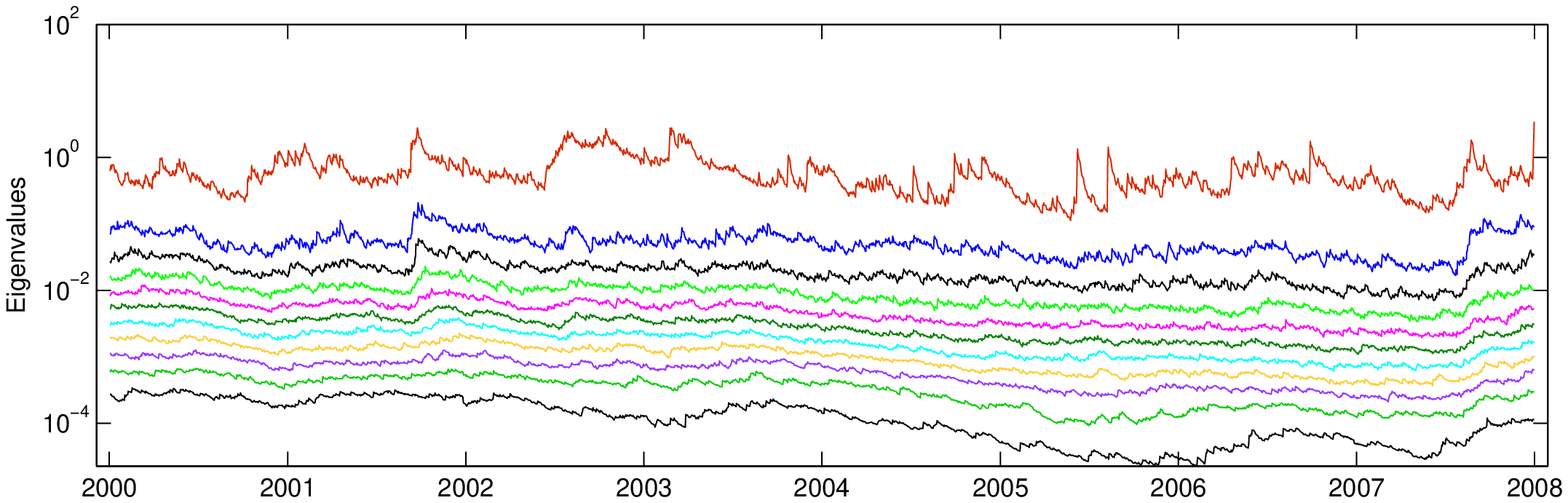}
  \includegraphics[width = 1.0\textwidth]{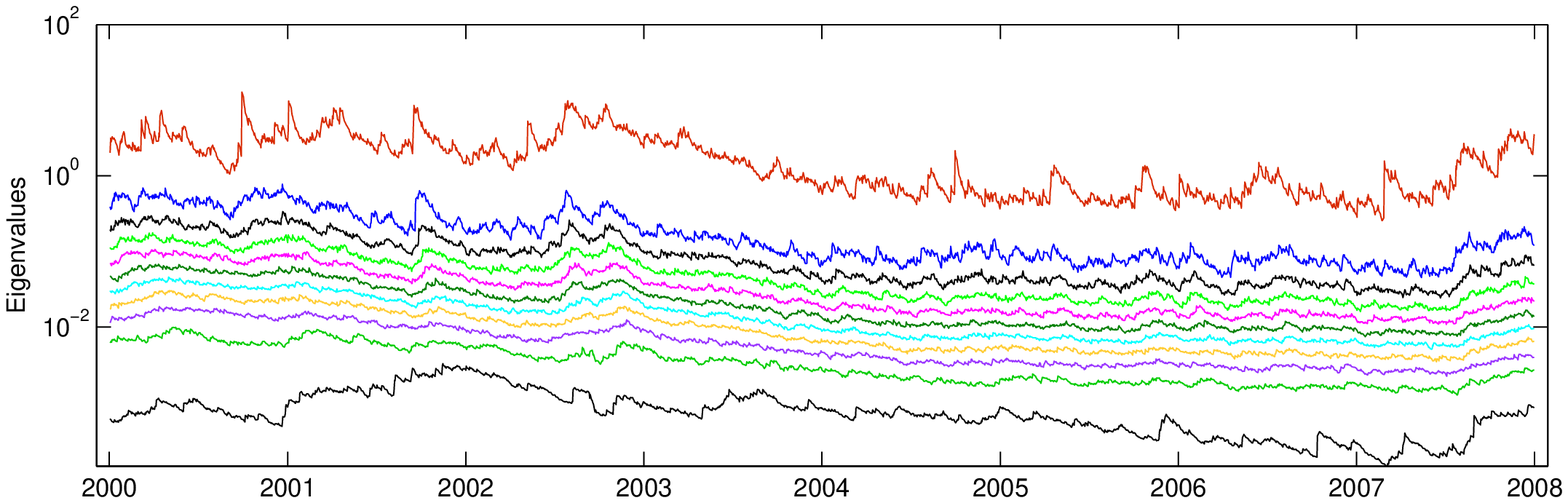}
  \caption{Eigenvalues of the daily covariance matrix. Each 10th eigenvalue  (i.e. $~\ev_{1}, \ev_{11}, \ev_{21}, \ev_{31},\ldots$) for the ICM dataset (top), each 5th eigenvalue for the G10 (middle) and USA (bottom) datasets}
  \label{fig:dynamicsalSpectrum_whole}
\end{figure}

\begin{figure}
  \centering
  \includegraphics[width = 1.0\textwidth]{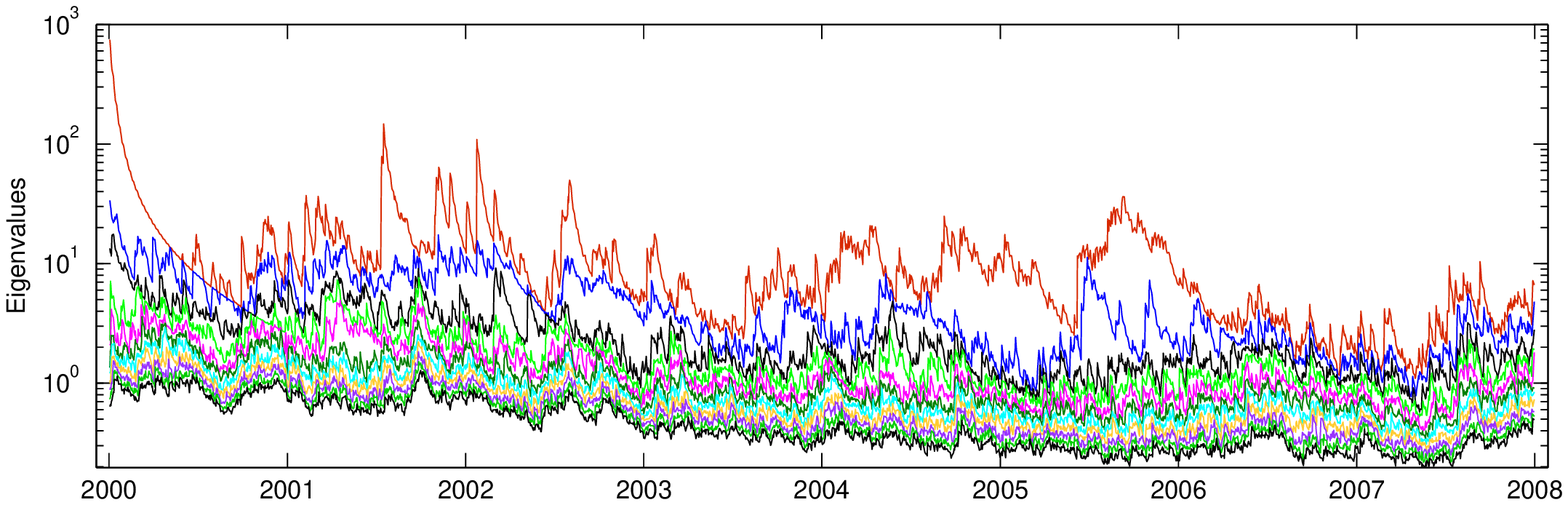}
  \includegraphics[width = 1.0\textwidth]{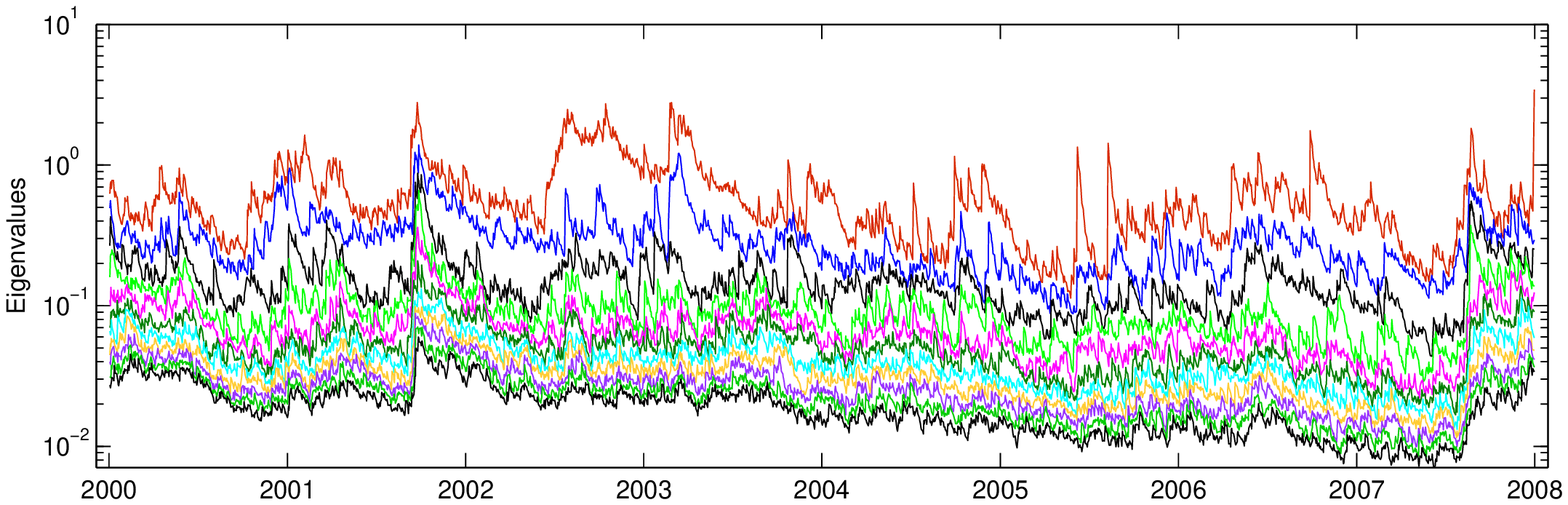}
  \includegraphics[width = 1.0\textwidth]{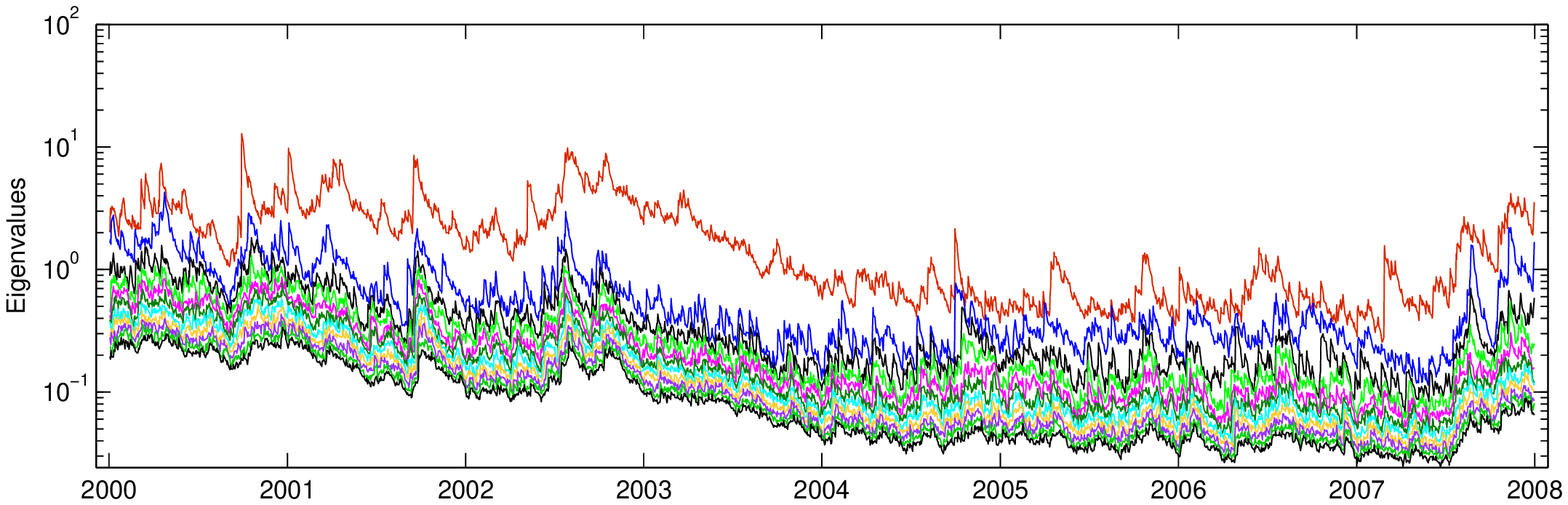}
  \caption{Largest eleven eigenvalues of the daily covariance matrix, for the ICM (top), G10 (middle) and USA (bottom) datasets.}
  \label{fig:dynamicsalSpectrum_top}
\end{figure}
The dynamics of the covariance spectrum are plotted in Figures~\ref{fig:dynamicsalSpectrum_whole} and \ref{fig:dynamicsalSpectrum_top}.
Figure~\ref{fig:dynamicsalSpectrum_whole} shows that the core of the spectrum is essentially independent of the time, except for a slow global dynamics measured for example by the mid eigenvalue $\ev_{N/2}(t)$.
The global slow dynamics has a time scale of a few years, while the peaks of volatility on the first eigenvalues have essentially no influence on the deep eigenvalues.
Only two events---namely the 11 September 2001 attack and the start of the subprime crisis (August 2007)---have had an influence on the core eigenvalues.
Interestingly, the largest eigenvalue of the ICM dataset shows no peculiar behavior on 11 September 2001, and is in fact related to the Argentina currency crisis in the same year.
The top ten eigenvalues have a richer structure, as seen on Figure~\ref{fig:dynamicsalSpectrum_top}.
The first five eigenvalues have a large and independent dynamics, while the next five eigenvalues move much more in synch.
Several eigenvalue crossings can also be observed, hinting at different subspaces with changing ranks.
The bottom line is that only a handful of eigenvalues have meaningful dynamics, and the rest can be described by a static distribution around a slow moving mean.

By analyzing the eigenvectors, the major peaks on Figure~\ref{fig:dynamicsalSpectrum_top} can be identified.
For the ICM dataset, they correspond to crises related to a single time series, and the stock indexes often do not belong to the first three eigenvectors.
For the G10 dataset, the first eigenvector often corresponds to the stock indexes, but the peaks are mostly related to natural gas and short-term EUR interest rates.
For the USA dataset, the stock and stock indexes make up most of the first eigenvector, with a few peaks created by single stocks or sectors.
The US interest rates appear only while the subprime crisis unfolds.
Overall, this analysis shows that the leading eigenvectors are often changing direction, and that it is not possible to select {\it a priori} stable leading directions.
% \begin{table}
% \centering
% \begin{tabular}{|l|l|}
% \hline
% Jan 2000  &  FX-Belarussia \\
% Aug 2001 &  IR-Argentina \\
% Feb 2004 & IR-Croatia \\
% Oct 2004 to May 2005 & IR-Croatia \\
% Jul 2005 & IR: Croatia, Russian, all IR  \\
% Oct 2005 & IR-Croatia \\
% \hline
% \end{tabular}
% \caption{Major peak for the first eigenvalue(s) of the ICM dataset.}
% \end{table}
% \begin{table}
% \centering
% \begin{tabular}{|l|l|}
% \hline
% Sep 2001		& All stock index + US interest rates \\
% Jul 2002 to Feb 2003   &  All stock indexes \\
% Mar 2003		& Natural Gas   \\
% Nov 2003		& IR-Euro 1d   \\
% Dec 2003		& Natural Gas   \\
% Nov 2004		& Natural Gas   \\
% Jun 2005		&  IR-Euro 1d   \\
% Sep 2005		&  IR-Euro 1d   \\
% Oct 2006		&  Natural Gas   \\
% Sep 2007		& US IR, All stock index, Natural Gas \\
% \hline
% \end{tabular}
% \caption{Major peak for the first eigenvalue(s) of the G10 dataset.}
% \end{table}
% \begin{table}
% \centering
% \begin{tabular}{|l|l|}
% \hline
% Oct 2000		& Apple +  technology stocks \\
% Oct 2004		& Merk \\
% May 2005		& All stocks \\
% Mar 2007		& All stocks \\
% Sep 2007		& 1d US interest rate  \\
% \hline
% \end{tabular}
% \caption{Major peak for the first eigenvalue(s) of the USA dataset.}
% \end{table}

The analysis of the leading eigenvectors and eigenvalues shows that they are dominated by particular events and crises.
On the other hand, the deeper eigenvalues only contain information about the overall volatility of the market corresponding to the selected investment space.
For example, the middle eigenvalue can be a good simple measure of the overall base market volatility.
In turn, this eigenvalue can serve as a baseline for the volatility, and the differences between the top eigenvalues and the baseline illuminate peculiar crises related to a few time series.

\FloatBarrier
\section{The dynamics of the correlation spectrum}
%-------------------------------------------------------------------------------------------
\label{sec:correlationSpectrum}
\begin{figure}
  \centering
  \includegraphics[width = 1.0\textwidth]{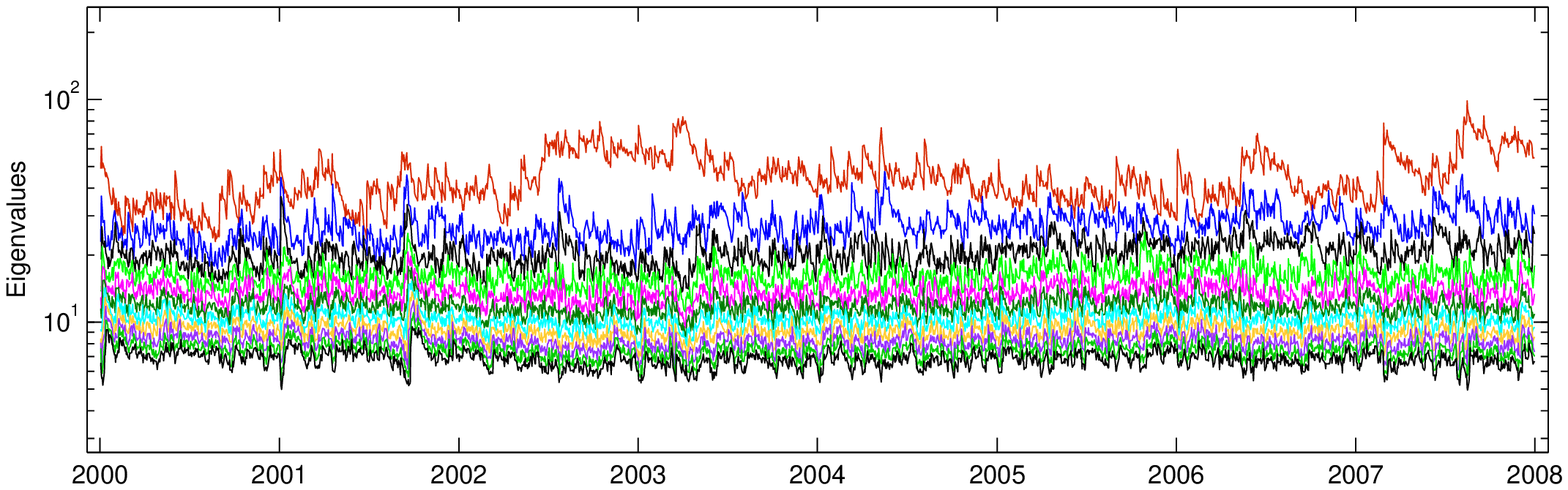}
  \includegraphics[width = 1.0\textwidth]{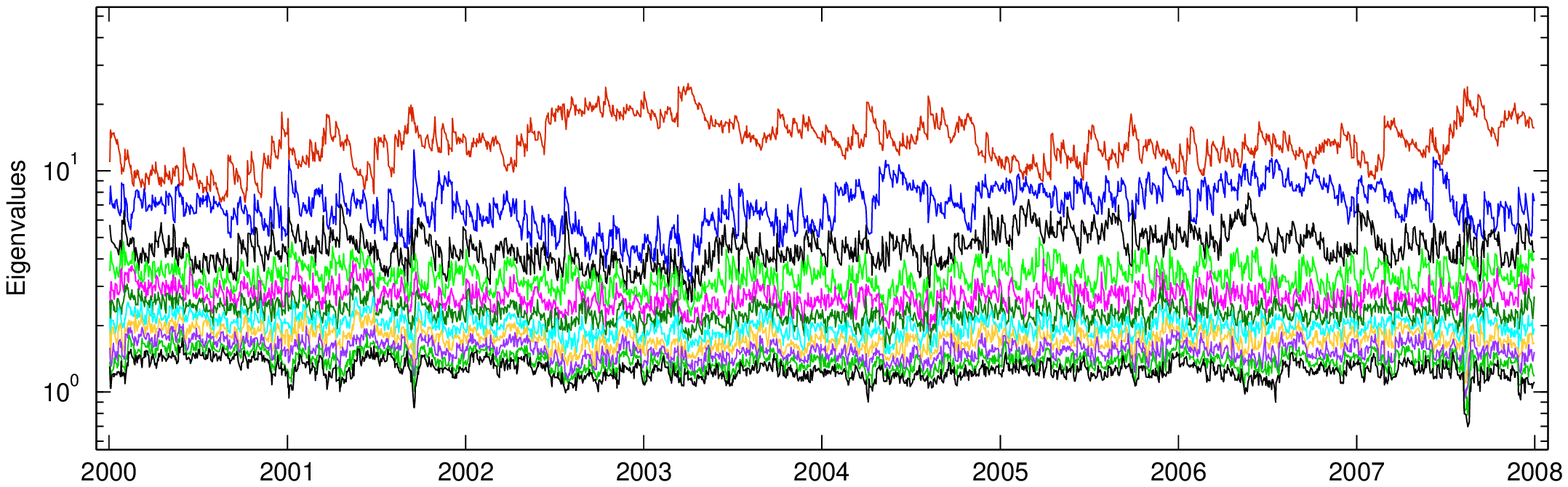}
  \includegraphics[width = 1.0\textwidth]{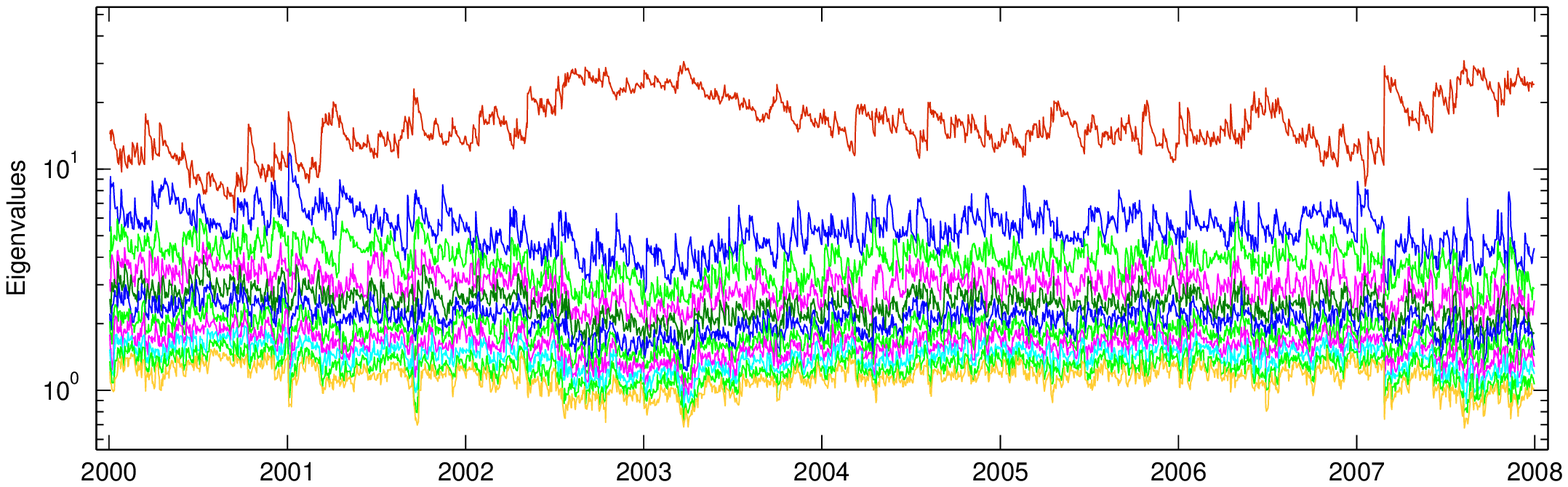}
  \caption{Largest eleven eigenvalues of the daily correlation matrix, for the ICM (top), G10 (middle) and USA (bottom) datasets. The upper vertical limits are $N_\text{pos}$, the largest possible values for the first eigenvalue, corresponding to perfectly correlated time series. The vertical span is a factor of 100 for all subgraphs.}
  \label{fig:dynamicsalSpectrum_top_correlation}
\end{figure}
The eigenvalues of the correlation matrix have to obey more constraints: in particular the spectrum is bounded by $0 \leq \ev_\alpha \leq N$ and $\sum_\alpha \ev_\alpha = N$. Moreover, the upper bound $\ev_1 = N$ for the first eigenvalue is reached for a perfectly correlated system with $\rho_{\alpha, \beta} = 1$.

The same dynamical analysis can be carried out for the spectrum of the correlation matrix.
Overall, the correlation spectrum is very static, in agreement with the common lore that the volatilities capture the largest part of the dynamics.
This is also consistent with the existence of a unique limit distribution of eigenvalues in large random correlation matrices.
Only the first three to five eigenvalues have an interesting dynamics, as reported in Figure~\ref{fig:dynamicsalSpectrum_top_correlation}.
%The characteristic time(s) for the fluctuations appears very similar to the characteristic time(s) for the covariance eigenvalues.

By comparing Figures~\ref{fig:dynamicsalSpectrum_top} and \ref{fig:dynamicsalSpectrum_top_correlation}, the similarity of the dynamics for the largest eigenvalues is clear.
The volatilities have larger moves, but the time scales involved in both graphs are similar, roughly of the order of one month.
This finding supports a joint evaluation of the covariance including the correlation and volatility, and against a model with a simple static behavior for the whole correlation matrix.

\FloatBarrier
\section{Spectral density of the correlation matrix}
%--------------------------------------------------------------------------------
\label{sec:spectralDensity_correlation}
The base description of financial time series is a Gaussian random walk, where the returns $r_\alpha(t)$ are iid Gaussian random variables.
The covariance elements are a simple sum of products of random variables (when using constant weights $\lambda(i) = 1/T$ and $T = \iMax$) with
\begin{equation}
 \Sigma_{\alpha, \beta} = \frac{1}{T} \sum_{1 \leq t \leq T} r_\alpha(t)\;r_\beta(t)\qquad\qquad\text{with } r_\alpha(t) \sim N(0,1).
\end{equation}
When the distribution of returns have a unit variance, the covariance matrix is equal (in average) to the correlation.
The spectrum of these {\it correlation} matrices has been  studied extensively, going back to the work of \cite{Wishart}.
The set of correlation matrices built from independent returns with unit variance is called the Wishart ensemble.
In this ensemble, under Gaussian returns, Marchenko and Pastur (1967)\nocite{MarcenkoPastur} derived the spectral density  of correlation matrices in the limit of large matrix size $N$ and large historical sample size $T$, for a fixed ratio $q = N/T$.
The M-P spectral density is given by
\begin{equation}
   \rho(\lambda) = \frac{\sqrt{4\lambda q - \left(\lambda + q - 1 \right)^2}}{2\pi\lambda q}
	\qquad\lambda\in\left[ (1-\sqrt{q})^2, (1+\sqrt{q})^2 \right].
\end{equation}
This spectrum has bounded support, and is generated purely by the random nature of the uncorrelated returns.
The idea here is to apply the theoretical upper bound of the M-P spectral density to empirical correlation matrices, in order to separate the significant spectrum (which is driven by true time series dependencies) from the noise-induced spectrum (which is described by the M-P density).

Unfortunately, the M-P spectral density is not precisely applicable to our case. First, the density is derived under the assumption that returns are Gaussian distributed, whereas empirical returns are best described by the Student distribution. Second, the M-P derivation assumes the correlation is computed from a constant weight scheme, whereas we use the long-memory weights. A last important difference is that the analytical computation is done in the limit $T\rightarrow\infty$, whereas the empirical investigation is done for a fixed kernel, corresponding to a constant $T$.
More precisely, the analytical computations are done in the limit $N\rightarrow\infty$, $T\rightarrow\infty$ with $q = N/T$ fixed, and the only parameters appearing in the final equations is $q$.
In fact, the present empirical investigation is better described by the limit $N\rightarrow\infty$, but with a fixed (long-memory) correlation kernel (the equivalent of $T$).
Unfortunately, there are no analytical results in this limit.
The theoretical computation of Marchenko and Pastur has been extended in several directions by \cite{PotterBouchaudLaloux}, leading to implicit equations for the spectral density, but no analytical result exists matching our precise needs.

\begin{figure}
  \centering
  \includegraphics[height = 0.4\textheight]{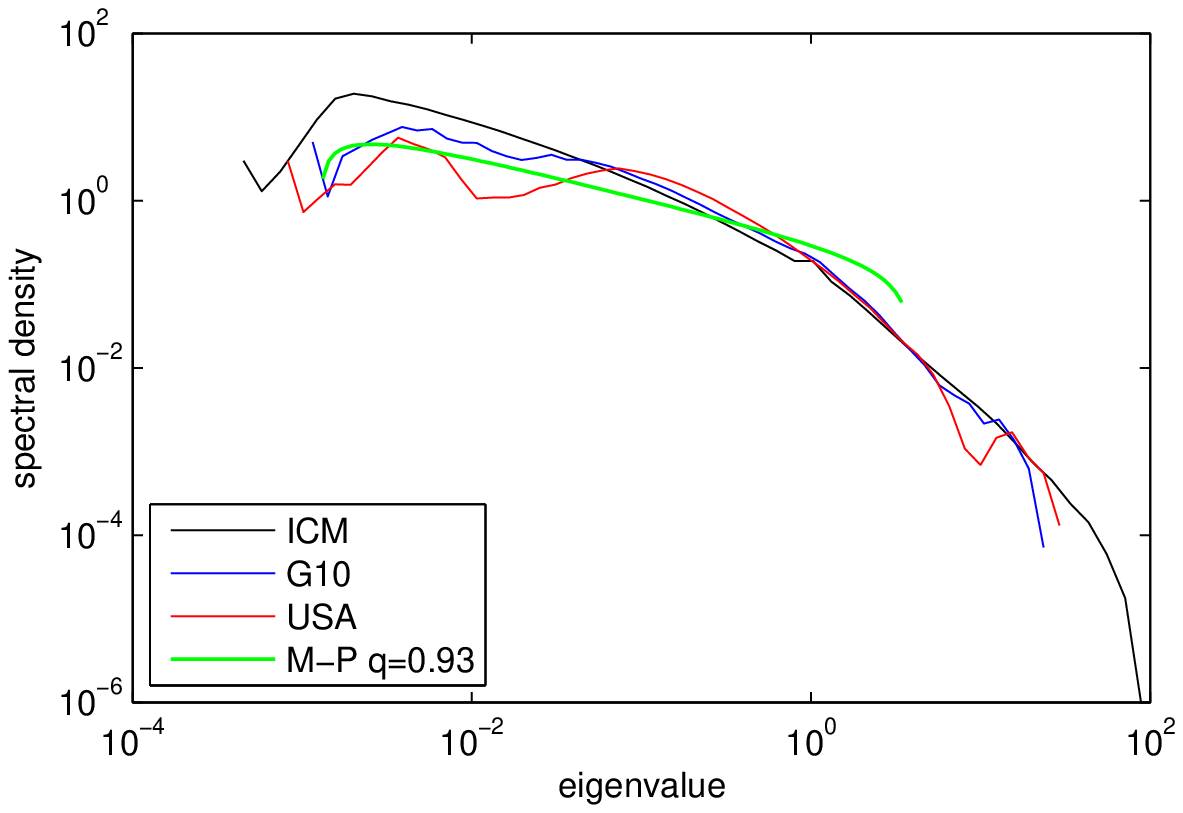}
  \caption{Mean spectral density of the correlation matrix $\rho$.
The green curve is the Marchenko-Pastur spectral density with $q = 0.93$ (chosen to provide the best overall fit) for a Gaussian-Wishart ensemble.}
  \label{fig:spectralDensity_correlation}
\end{figure}
The empirical mean spectral density can be computed from the time evolution of the spectrum
(in effect, replacing an ensemble average in the theoretical computation by the time average of the empirical spectrum).
The spectral density of the correlation matrix is shown in Figure~\ref{fig:spectralDensity_correlation} for the three datasets.
The similarity between the three datasets is striking, particularly since the ratio $q = N/T$ varies by a factor of six across the datasets.
Moreover, there is a lack of any feature that would separate the noise spectrum from the significant spectrum. 
And whereas the deviation between the M-P and empirical density in the upper part of the spectrum can be attributed to correlations in the market, the M-P density also does not describe the empirical data in the lower part of the spectrum.
This can be explained by the discrepancies mentioned above.

For our purposes, we are interested mostly in {\it covariance} matrices.
The key difference is the volatility, which has an important dynamic and serial correlations.
For the covariance (with an assumption about the distribution for the volatilities), there seems to be no known analytical results about the spectral density.
However, we can expect to have the most important eigenmodes related to the volatility dynamics, first of the market, then of the most important market factors.
We can also expect to have a similar dense part for the lower section of the spectrum, corresponding essentially to random noise.

\section{Spectrum and spectral density of the covariance}
\label{sec:spectrum_covariance}
%------------------------------------------------------------
\begin{figure}
  \centering
  \includegraphics[height = 0.39\textheight]{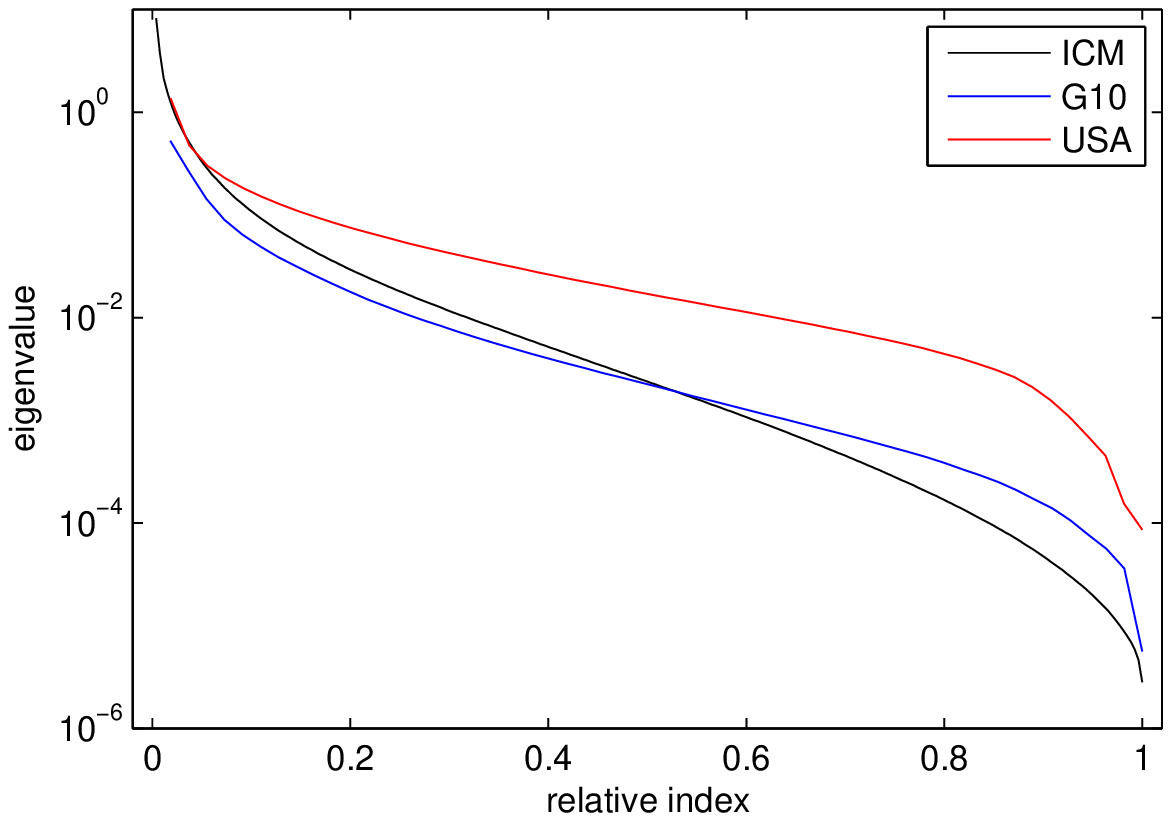}%original scaling [width = 0.4\textwidth]
  \caption{Mean spectrum of the covariance matrix $\SigmaEff$ plotted as a function of the relative index $\alpha' = \alpha/N_\text{pos}$}
  \label{fig:spectrumSigma_mean}
\end{figure}
The mean spectrums of the covariance for the three datasets are plotted on Figure~\ref{fig:spectrumSigma_mean}.
Despite the fairly different sizes and universes, their overall shapes are very similar.
They are well described by the simple Ansatz
\begin{eqnarray} \label{eq:AnsatzEpsilonVsAlpha}
 x & = &\frac{1}{2} ~-~ \frac{\alpha}{N},  \qquad\qquad \alpha = 1, \cdots, N \nonumber\\
 \ln(\ev_\alpha) & = &  \ln(\ev_{N/2}) + ax/\left[1 - \left(\frac{2x}{b}\right)^4\right].
% \ln(\ev) =   \ln(\ev_{N/2}) + \frac{a}{b}\sinh\left\{ b\left(\frac{1}{2}-\frac{\alpha}{N}\right)\right\}
% \ln(\ev_\alpha) =   \frac{1}{N-1}\left\{ (N-\alpha)\ln(\ev_\text{max}) + (\alpha - 1)\ln(\ev_\text{min}) \right\}
\end{eqnarray}
The constant $a$ controls the slope at the center of the spectrum, and $b$ the quartic curvature on both ends.
This Ansatz indicates that the core of the spectrum decays exponentially fast toward zero
\begin{equation}
 \ev_\alpha \simeq \exp\left(-\frac{a\alpha}{N}\right),
\end{equation}
potentially leading to troubles when computing the inverse of the covariance.
% The parameters used on the Figure~\ref{fig:spectrumSigma_mean} are
% \begin{center}
% \begin{tabular}{l|ll}
%            & $a$ (slope) & $b$ (curvature)\\
% \hline
%  ICM  & 8.0 & 1.35 \\
%  G10  &  5.8 & 1.33 \\
%  USA  & 4.3 &  1.16
% \end{tabular}
% \end{center}
% Other Ansatze like $ \ln(\ev_\alpha) \simeq a/b~\sinh(bx)$ also provide for a good spectrum fit,
% but the quadratic curvature close to both ends is too weak.

Up to a normalization by $N$, the index $\alpha$ is essentially the cumulative density of state, with
\begin{equation}
   \alpha = \alpha(\ev) = N\int_{\ev}^{\ev_\text{max}} d\ev' \; \rho(\ev'),
\end{equation}
and where $\ev_\text{max}$ is the largest eigenvalue.
The density of states is obtained by derivation of the above equation
\begin{equation}
 \rho(\ev) = -\;\frac{1}{N}\;\frac{\partial\alpha}{\partial\ev}
    ~=~  -\;\frac{1}{\ev}  \;\frac{1}{N\;\partial\ln\ev/\partial\alpha}.
\end{equation}
The Ansatz for the spectrum leads to a scale free density of states
\begin{eqnarray}
\label{eq:densityOfStates}
  \rho(\ev) & = & \frac{1}{a~\ev} + \text{curvature correction}.
\end{eqnarray}
%This density of state likely corresponds to the part of the spectrum which, in the large size limit, is continuous (after the probability distributions for the volatilities would have been suitably defined).
%The leading $1/\ev$ density provides for a scale free density, while the next term gives the curvature correction close to the spectrum  hedge.
Notice that this shape is quite different from the mean spectrum of correlation matrices.

%\FloatBarrier
%\section{Spectral density of the covariance}
%--------------------------------------------------------------------------------
%\label{sec:spectralDensity_covariance}
\begin{figure}
  \centering
  \includegraphics[height = 0.39\textheight]{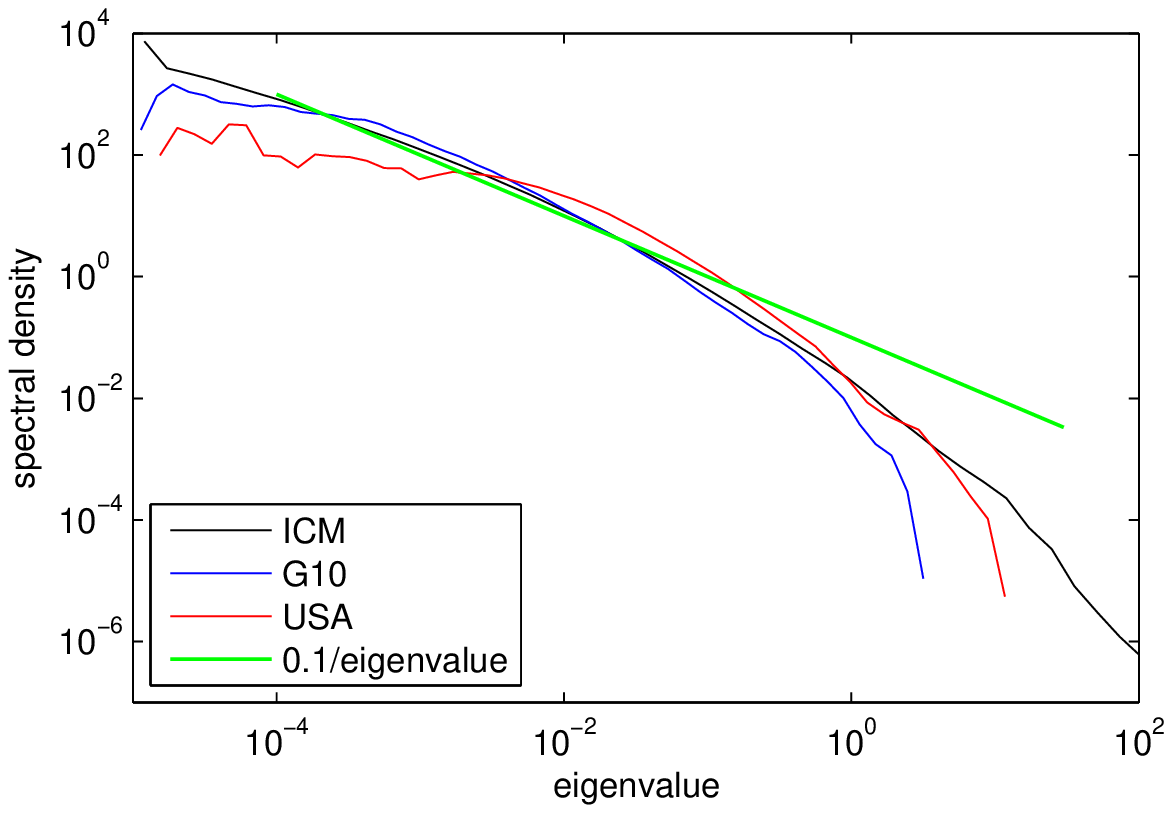}%original scaling [width = 0.4\textwidth]
  \caption{Mean spectral density of the covariance matrix $\SigmaEff$.
The green line corresponds to $\rho(\varepsilon) = 0.1/\varepsilon$.}
  \label{fig:spectralDensity_covariance}
\end{figure}
Similarly to the spectral density for the correlation, the spectral density for the covariance is given in Figure~\ref{fig:spectralDensity_covariance}.
The three densities are very similar (with a displacement toward higher eigenvalues for the USA dataset because it contains mostly stocks that are more volatile).
The green line corresponds to $\rho(\varepsilon) = 0.1/\varepsilon$ as in \eqref{eq:densityOfStates}, equivalent to a linear shape for the logarithmic spectrum.
This very simple form describes well the empirical density of states over two to three orders of magnitude,
while the upper part of the densities clearly decays faster.
Notice that there is again no special feature that would separate the leading significant eigenvalues from a bulk density originating in random matrices.

\FloatBarrier
\section{Mean projector}
%----------------------------------------------------
\label{sec:meanProjector}

\begin{figure}
  \centering
  \includegraphics[width = 0.75\textwidth]{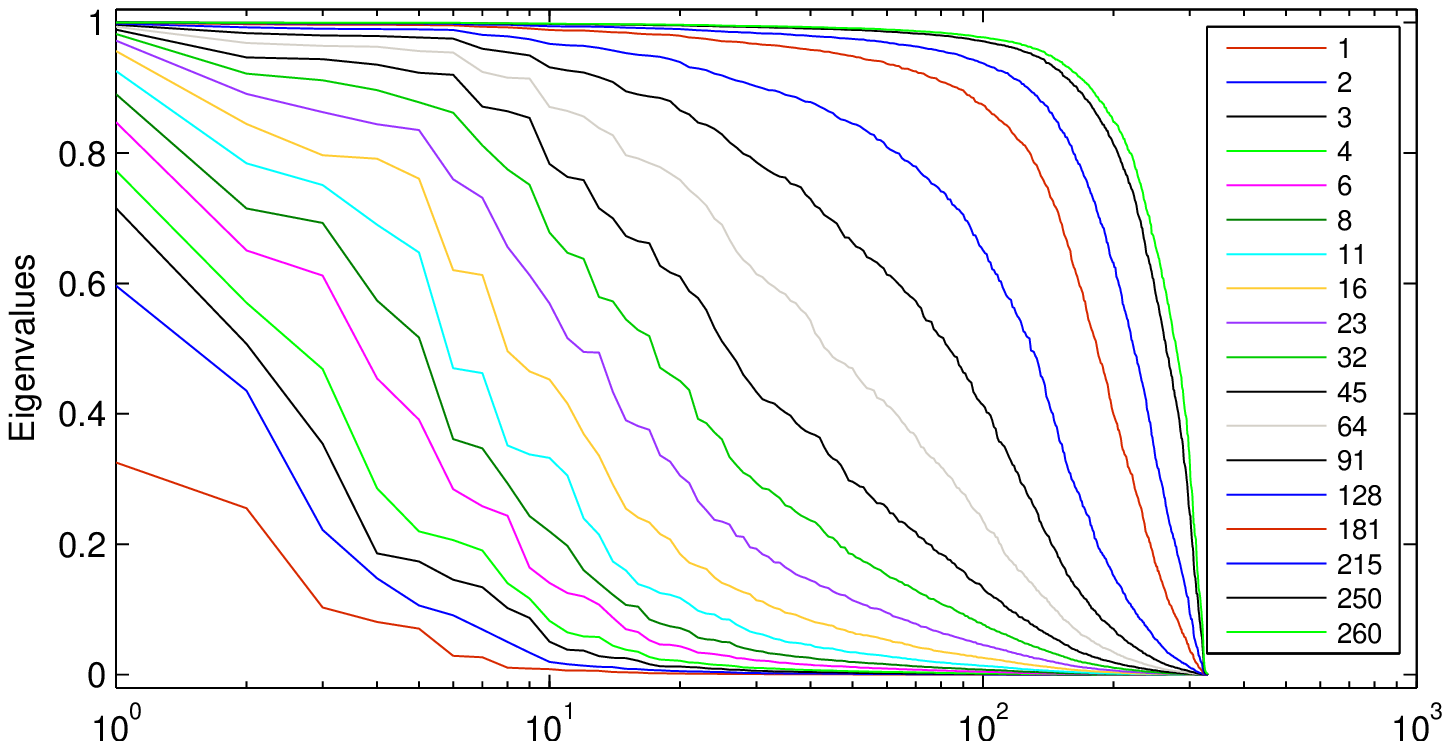}\\ %original scaling [width = 0.8\textwidth]
  \includegraphics[width = 0.75\textwidth]{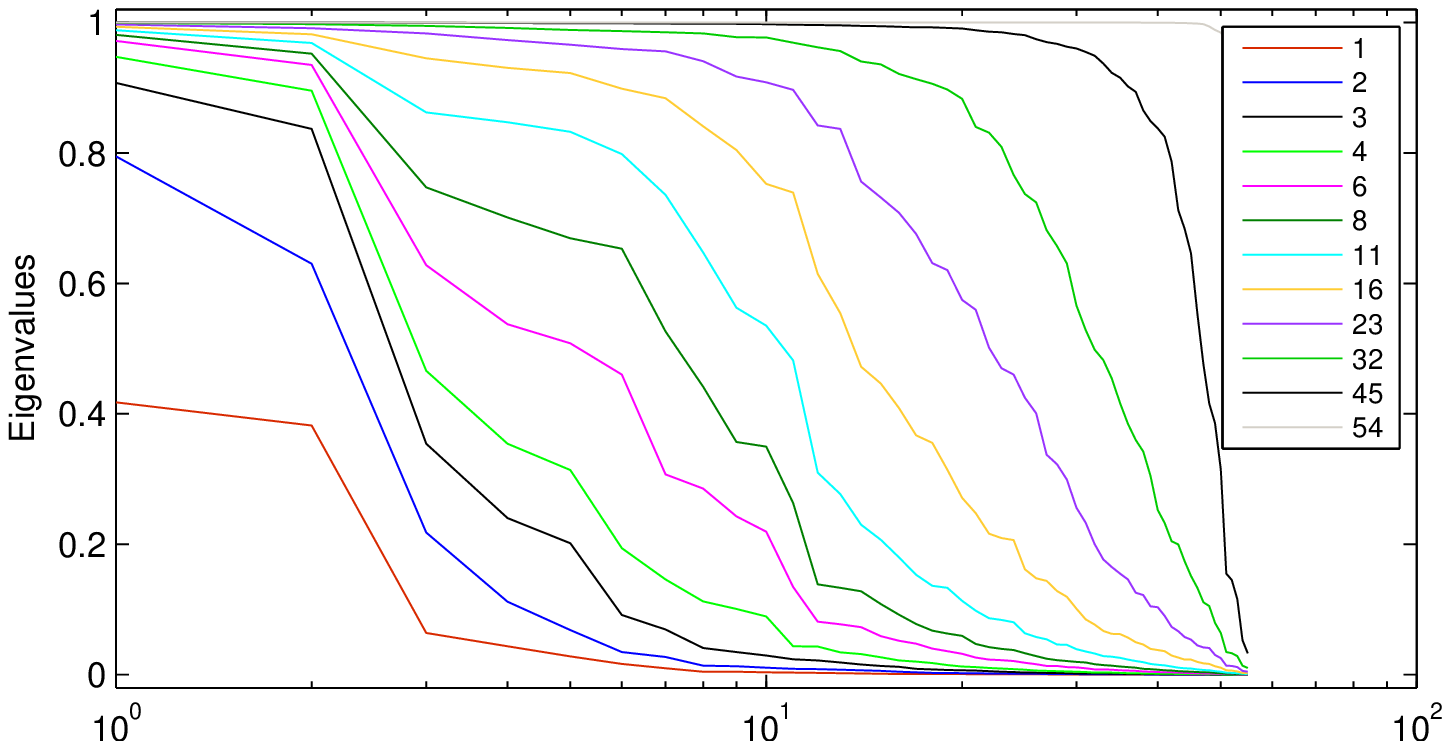}\\
  \includegraphics[width = 0.75\textwidth]{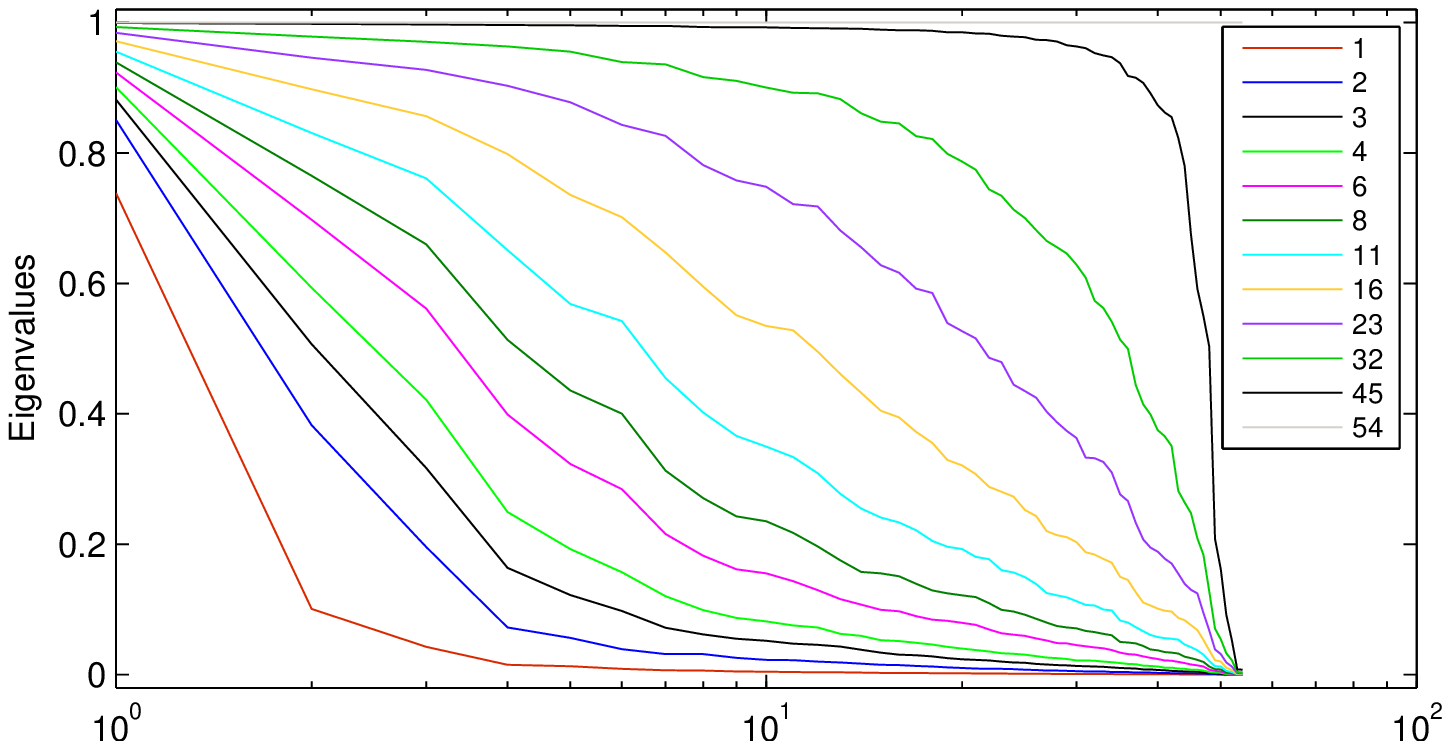}
  \caption{Spectrum of the mean projector as function of the projector rank, for the  ICM (top), G10 (middle) and USA (bottom) datasets.}
  \label{fig:SpectrumMeanProj_FixedRank}
\end{figure}
The spectrum of the covariance gives only a very partial view of the dynamical behavior of a matrix time series.
In particular, the directions and subspaces associated with the eigenvalues are important.
In order to gain an insight into the leading subspace behavior, the mean projector is defined in \eqref{eq:meanProjector}.
Figure~\ref{fig:SpectrumMeanProj_FixedRank} shows the spectrum of the mean projector for increasing projector rank for the ICM, G10, and USA datasets.
As the rank of the projector increases, the mean projector changes from the null matrix to the identity matrix.
For the intermediate ranks, the eigenvalues decrease gradually, without gaps in the spectrum.
Only cusps can be seen in the spectrum behavior with increasing rank.
For the ICM dataset, cusps can be observed around ranks two and seven, with the smaller rank eigenvalues being close to one, and the larger rank eigenvalues decreasing  rapidly.
For the G10 dataset, a change of behavior can be observed around ranks two and six.
For the USA universe, which is dominated by stocks, the leading eigenvector is clearly dominant for small ranks, and a weak cusp can be observed around ranks three to four.
For the G10 and USA, the first eigenvalue is dominant for small ranks, but less so for the ICM dataset.
Globally, the image is more of a cascade of subspaces with regularly decreasing importance, while the leading eigenvalue (corresponding to the market) is dominant for small ranks.
For the three datasets, the ranks for the cusps are in line with the complexity of the datasets, namely for a richer universe, a larger number of important directions is needed.
This is also in line with Figure~\ref{fig:dynamicsalSpectrum_top}, where the USA dataset has the smaller number of eigenvalues with dynamics different from the bulk.

The coordinates of the eigenvectors corresponding to the first eigenvalues (of the mean projector) can be analyzed.
These measure the direction corresponding to the leading subspace, and they can be interpreted as loadings in a (mean) PCA decomposition.
For the G10 dataset, the first eigenvector has a weight of one on the gas future contract and zero for all other time series. The second eigenvector corresponds to a fairly complex strategy: essentially long FX and commodities, short stocks and interest rates.

For the USA dataset, the first eigenvector shows positive weights for all stocks and stock indices and clearly is associated with the overall stock market. The second vector has mainly negative weights for high-tech
stocks and positive weights for energy stocks, while other traditional stocks have negligible weights.
The third eigenvector is essentially long Apple with weight $\simeq 0.9$, and short most of the other stocks; the fourth eigenvector corresponds to a complex long/short strategy on stocks.
On all four eigenvectors, the interest rates have a negligible weights.

The first ICM eigenvector is associated with the Russian interbank interest rate at one day;
the second eigenvector is essentially a long position on the short term Croatian interbank rate (with a weight of one) with small long positions on stocks and short positions on short term interest rates; the third eigenvector is essentially a long position on stocks and stock indexes, with smaller long positions in commodities; the fourth eigenvector is short on Slovakia one-day deposit (with a weight of -0.6) and long on most other short-term interest rates. On this set, the third eigenvector corresponds to the common view of the ``global stock market volatility'', while the other main vectors are related to emerging markets.
These factors correspond to key drivers of the volatility but they might not have been chosen {\it a priori} by market analysts.
Notice that this computation is sensitive only to the selected universe, but not to a portfolio or to the economic weights related to a time series.

In conclusion, the properties of the mean projector spectrum show that it is not possible to identify clearly a dominant stable leading subspace.
Instead, the picture is of subspaces with regularly decreasing importance, without a threshold that would separate a significant subspace from a random subspace.
Even though the covariance spectrum and correlation spectrum are very static (except for a handful of top eigenvalues), the eigenvectors of the covariance have important dynamics deep in the spectrum.
This goes somewhat against the PCA picture \cite{Tsay-05} which explains the dominant stock dynamics by a few stable eigenmodes.
Similarly, the picture conveyed in factor models, as initiated by \cite{Fama-93}, is an oversimplification of the complex dynamic occurring between the time series.
Moreover, the leading directions occurring in large datasets would certainly not have been chosen {\it a priori} by most analysts.

%\FloatBarrier
\section{Projector dynamics}
%--------------------------------------------------------
\label{sec:projectorDynamics}
In order to have a better understanding of the dynamics of the subspaces associated with the projectors $\sfP_k(t)$, a fluctuation index and a scalar lagged correlation are defined.
The fluctuation index $\gamma$ essentially measures the difference between $\tr\avg{\sfP^2}$ and $\tr\left( \avg{\sfP}^2 \right)$, with a convenient normalization
\begin{equation}
 \gamma = \frac{\tr\avg{\sfP^2} - \tr\left( \avg{\sfP}^2 \right)}{\tr\left(\avg{\sfP}\right)}
	~=~ 1 - \frac{1}{k}\;\tr\left( \avg{\sfP}^2 \right).
\end{equation}
If the projector is mostly constant $\avg{\sfP} \simeq \sfP(t)$, then $\avg{\sfP}^2 \simeq \avg{\sfP}$ and the fluctuation index is close to zero.
In the other direction, if the projector dynamics explores fully the available space $\avg{\sfP_k} \simeq k/N \;\id$, and the fluctuation index reaches its maximal value $\gamma_\text{max} = 1 - k/N$.
The fluctuation index can be computed for increasing ranks $k$, and for projectors associated with the covariance or correlation matrices.

\begin{figure}
  \centering
  \includegraphics[width = 0.55\textwidth]{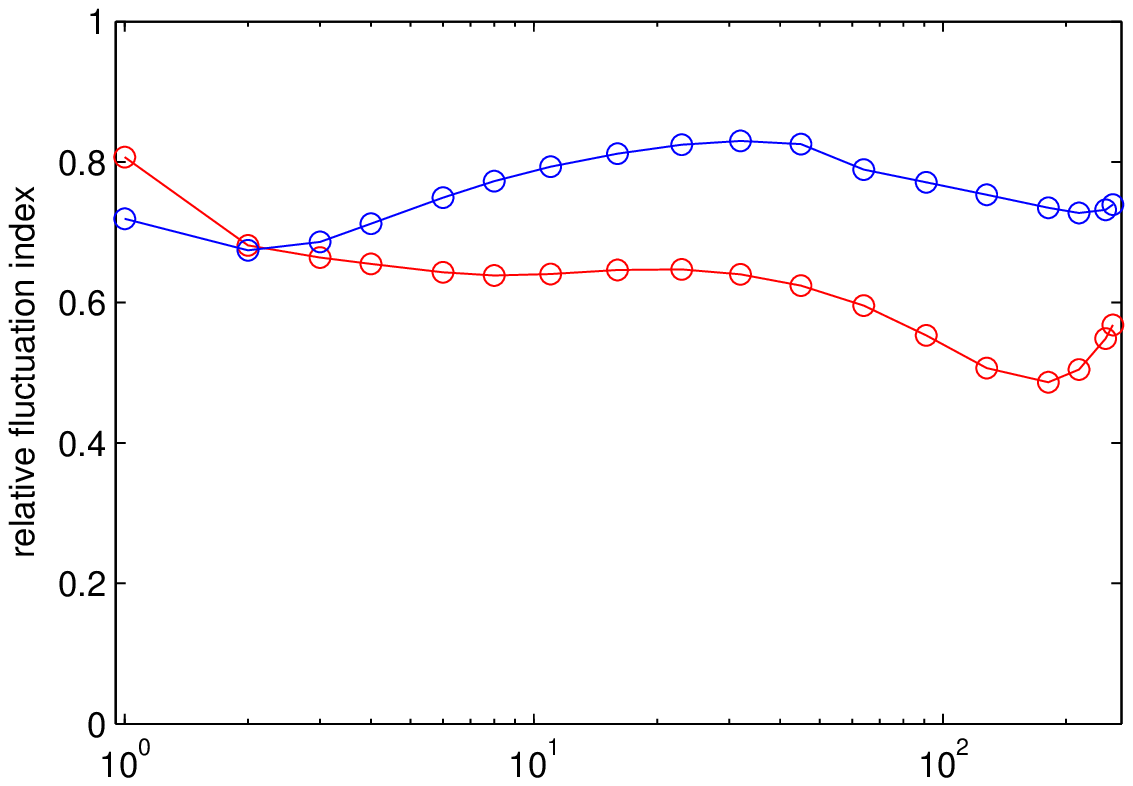}\\
  %\thisfigspace
  \includegraphics[width = 0.55\textwidth]{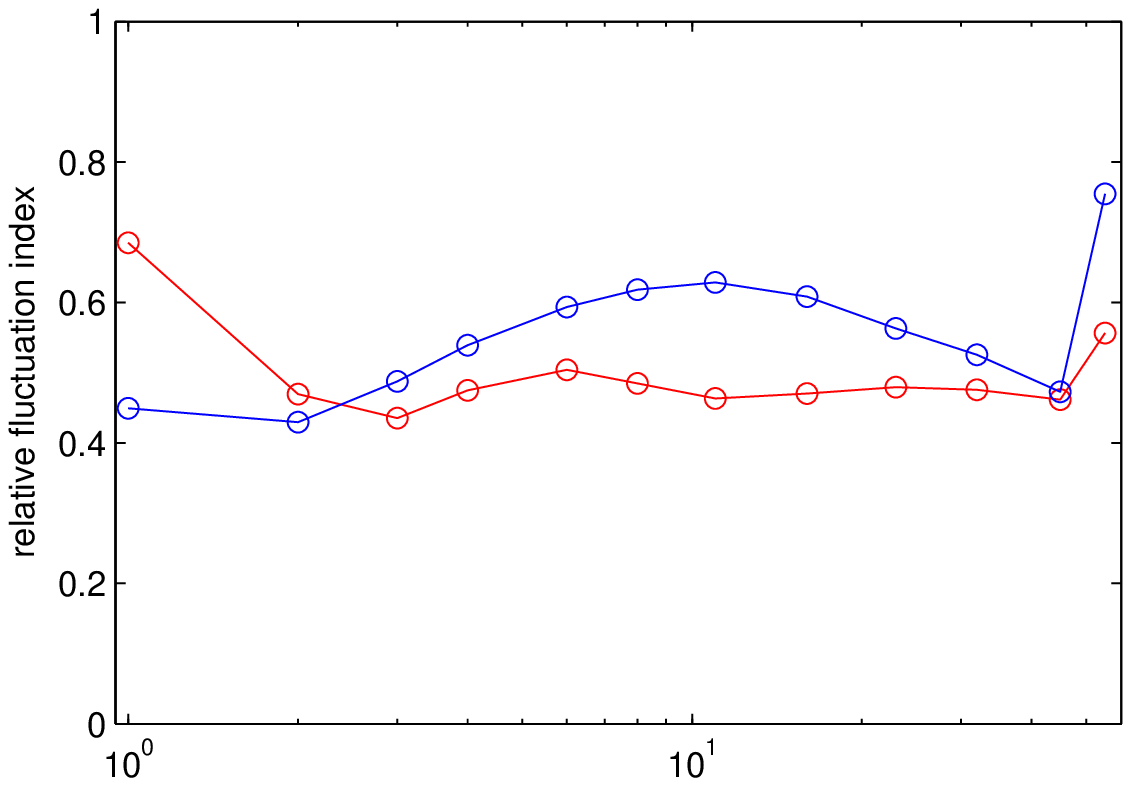}\\
  %\thisfigspace
  \includegraphics[width = 0.55\textwidth]{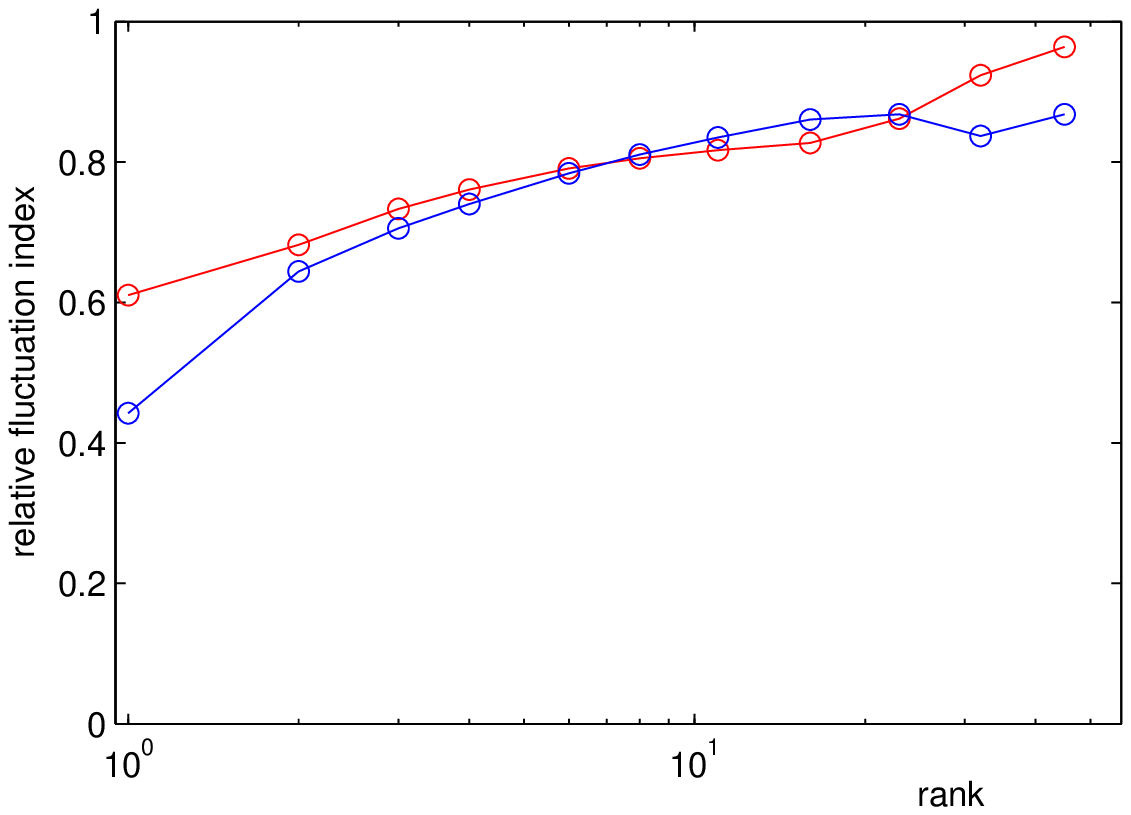}
  \caption{Relative fluctuation index $\gamma/\gamma_\text{max}$ as function of the projector rank, for the ICM (top), G10 (middle) and USA (bottom). Projectors of covariance (red) and correlation (blue). %The red (blue) curve is for the projectors computed from the covariance (correlation).
   }
  \label{fig:fluctuationIndex_projector}
\end{figure}
Figure~\ref{fig:fluctuationIndex_projector} displays the relative fluctuation index $\gamma/\gamma_\text{max}$ for the three datasets as function of the projector rank.
The fluctuation indexes are at least of the order of 40\% of the maximal value, and typically larger.
Moreover, the behavior is similar for the covariance and correlation.
This clearly shows that the eigenvectors always have an important dynamic behavior, although the eigenvalues can be fairly static.
Interestingly, the relative fluctuation indexes show a minimum for small sizes, between rank one for the USA dataset to rank eight for the covariance on the ICM dataset.
Another feature is the smaller fluctuation index for the projectors derived from correlation matrix that for the covariance matrix, for rank one and two.
These lower values show that the two first eigenvectors of the correlation are more stable than for the covariance.
These are the only indication in the present work of some increased stability for selected small ranks.

In order to assess more quantitatively the dynamics, we study the lagged correlations for different matrices (covariance, correlation, and the related projectors with increasing ranks).
In order to avoid correlations created by the long-memory weights, a compact kernel for $\lambda(i)$ is used for this particular computation when computing $\SigmaEff$: the weights $\lambda(i)$ are constant with $\iMax = 21$.
The value $\iMax = 21$ allows us to measure lagged correlations at shorter lags and with a larger effective sample size, but is rather low to obtain reliable estimates of the correlation matrix.
The results have been checked with $\iMax = 42$, without significant differences.
For a time series of matrices $\mathsf{X}(t)$, the scalar lagged correlation is defined by
\begin{equation}
 \rho(\tau) = \frac{\tr\avg{ \left(\mathsf{X}(t) - \avg{\mathsf{X}}\right)\left(\mathsf{X}(t+\tau) - \avg{\mathsf{X}}\right)}}
		{\tr\avg{ \left(\mathsf{X} - \avg{\mathsf{X}}\right)^2}}.
\end{equation}
Essentially, $\rho(\tau)$ quantifies the overall linear dependency between the matrices $\mathsf{X}(t)$ and $\mathsf{X}(t+\tau)$.

\begin{figure}
  \centering
  \includegraphics[width = 0.6\textwidth]{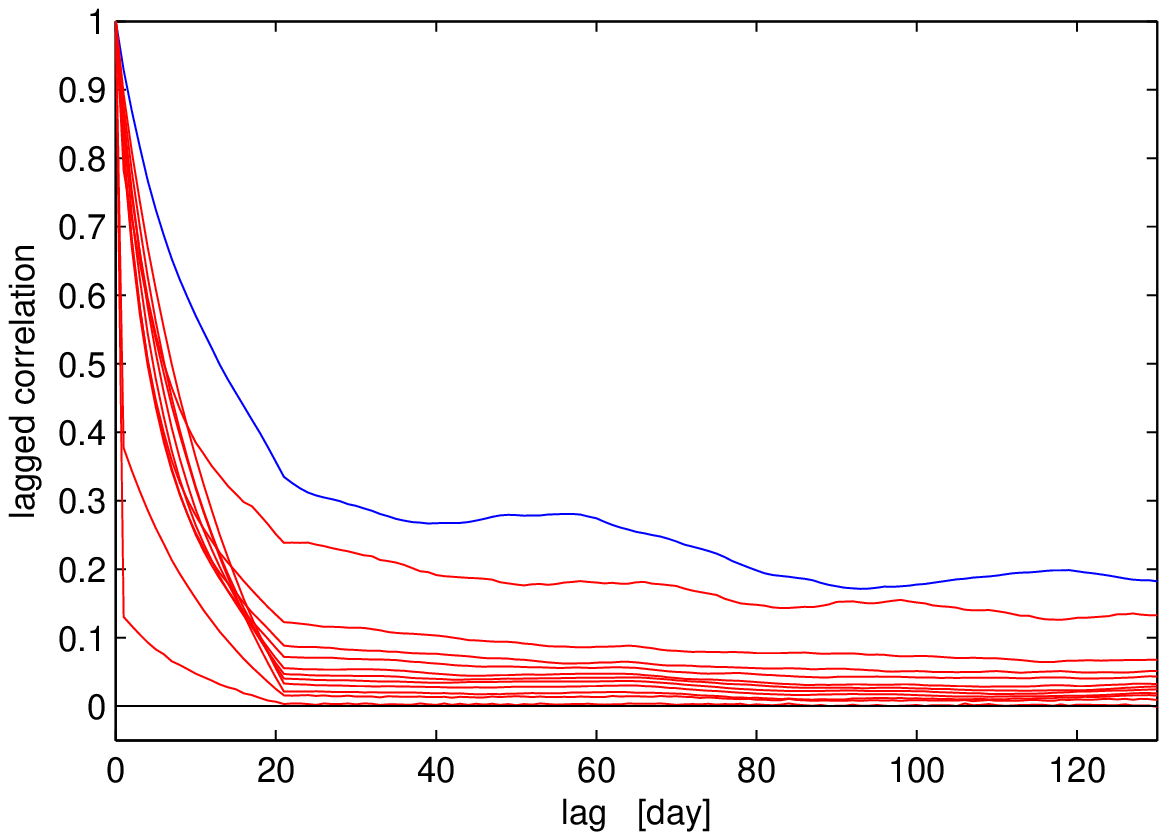}\\
  \includegraphics[width = 0.6\textwidth]{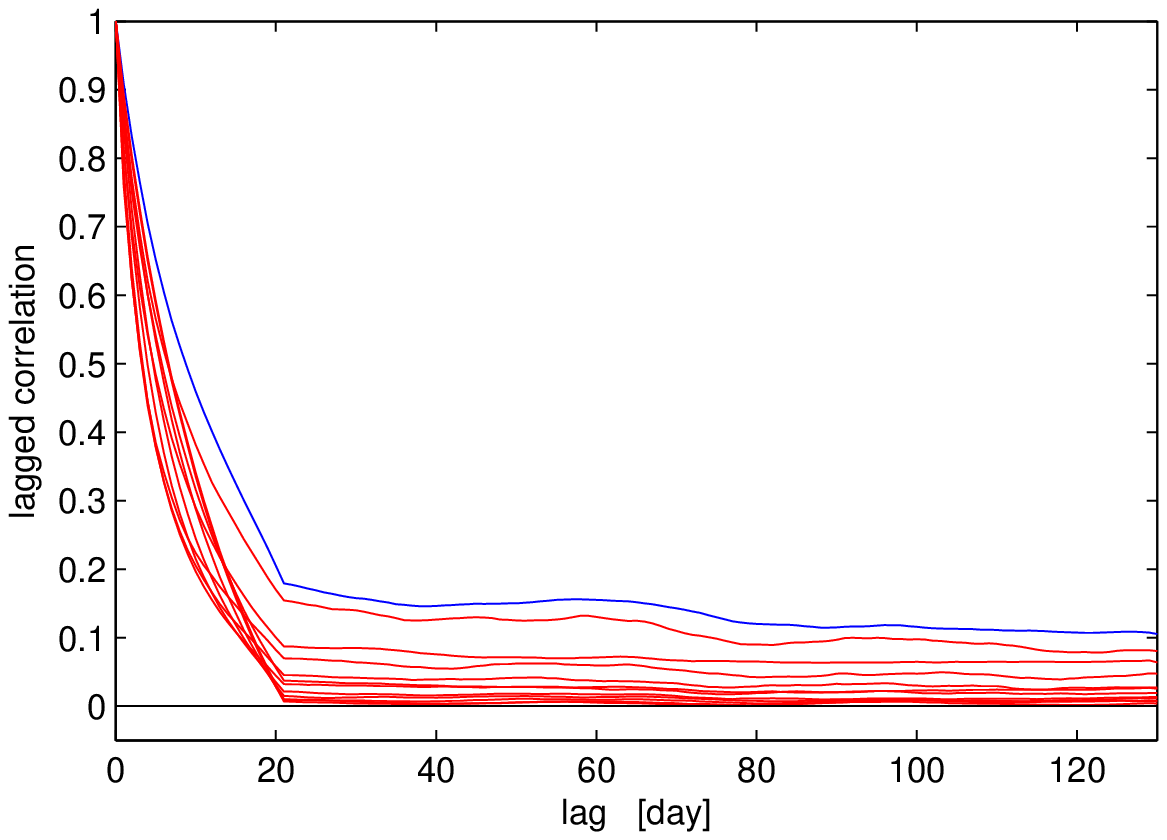}
  \caption{Lagged correlations for the USA dataset, for the Covariance (top) and correlation (bottom).  Results for full matrix (blue) and the projectors of increasing size (red, curves lying lower for increasing ranks).}
  \label{fig:laggedCorrelation}
\end{figure}
Figure~\ref{fig:laggedCorrelation} displays the lagged correlations for the USA dataset; the figures are similar for the ICM and G10 datasets.
For all curves, the regular drop between lags 1 to 21 is simply due to the overlap of the volatility estimated with 21 historical points.
The lag correlation for the covariance (top graph, blue curve) shows the long memory of the volatility, as visualized by the slow decay of the lagged correlation and the still high correlation level ($\simeq$20\%) after six months.
The corresponding projectors (red curves) have also a long memory, but with a regularly decreasing intensity for increasing ranks.
This shows that the first eigenvectors have also a long-memory dependency, but the lagged dependency is getting weaker for increasing rank.
The bottom graph shows the same quantities, but for the correlation.
The curves have the same qualitative behaviors, but with a lower overall level for the correlations.
The lower overall level can be expected as part of the lagged dependency in the covariance is due to the (single component) volatilities.
This clearly shows that the covariance and correlation matrices have the same qualitative behavior:
both have important long-memory dynamics, mainly for the eigenvectors.

\FloatBarrier
\section{Conclusion}
%-----------------------------------
Bringing together the pieces above, we see a better picture of the dynamic behavior of the covariance and correlation matrices emerging.
The eigenvalues of the covariance are mostly uninformative, except the first three to ten eigenvalues.
Essentially, only a handful of eigenvalues show meaningful dynamics, while the bulk of the spectrum is very static and well described by a simple distribution.
The transition from significant to noisy eigenvalues is gradual, and the spectrum and spectral density show no particular features that would separate them.
On the other hand, the corresponding eigenvectors have informative dynamics much deeper in the spectrum, without clear invariant subspaces corresponding to the main market modes.
Overall, the resulting image is of dynamic subspaces with regularly decreasing importance across the whole spectrum.

The dynamics of the correlation eigenvalues are even more uniform, indicating that the volatility dynamics are dominant over the correlation.
Yet the top three to five eigenvalues have a clear and distinct time evolution, and an approximation by a constant spectrum seems inappropriate.
The corresponding projectors have a very similar behavior compared to the projectors derived from the covariance, with an important dynamics.
Although random matrix theory can shed some light on the difference between significant and noise-induced eigenvalues, the important information contained in the eigenvectors stays hidden to such an approach.
Beside, the time scales for the correlation fluctuations are very similar from the time scales in the covariance fluctuations,
as both exhibit the signature of long memory as measured by the slow decay of the lagged correlations.
These similar properties show that a simple covariance estimator measures correctly both volatilities and correlations, and separate estimators are unnecessary.
For the mathematical description of multivariate time series, this points to a multivariate GARCH structure, and not to the separation of volatilities and correlations as initiated by \cite{Bollerslev.1990} with the CCC-GARCH process.

The definition of the covariance matrix that is investigated in depth corresponds to weights that decay logarithmically slowly, in order to
build a good one step volatility forecast based on the long memory for the volatility observed in financial time series.
Other shapes can be used for the weights $\lambda(i)$, including the ``equal weight in a window'' prescription, or exponential weights $\lambda(i) = (1-\mu)\mu^i$.
Subsequently, the spectrum of the covariance is dependent on the weights $\lambda(i)$.
Yet, there is nothing specific to the long-memory kernel, and a uniform or an exponential kernel shows essentially the same properties, with a spectrum that decreases exponentially fast toward zero (the pace for the decay is depending on the kernel), and with an important dynamics for the projectors.
This means that these key properties are generic, and only their quantitative aspects are influenced by the weights $\lambda(i)$.

The large number of small eigenvalues implies that the inverse covariance is ill defined.
In the computations that require the inverse covariance, an appropriate regularization should be used, even when the covariance is mathematically not singular.
The present investigation shows that no simple threshold can be defined, and that a large part of the information lies in the dynamics of the eigenvectors, both for the covariance and correlation.
The detailed analysis of the regularization schemes and their impacts on process inference will be investigated in a forthcoming paper \cite{Zumbach.inferenceOnProcesses}.
It can already be anticipated that a simple exclusion of the smallest eigenvalues, and corresponding eigenvectors, leads to poor results.

\bibliographystyle{apalike}
\bibliography{bibliography}

\begin{thebibliography}{}

\bibitem[Bollerslev, 1990]{Bollerslev.1990}
Bollerslev, T. (1990).
\newblock Modelling the coherence in short-run nominal exchange rates: A
  multivariate generalized {ARCH} model.
\newblock {\em Review of Economics and Statistics}, 72:498--505.

\bibitem[Briner and Connor, 2008]{BrinerConnor.2008}
Briner, B.~G. and Connor, G. (2008).
\newblock How much structure is best? {A} comparison of market model, factor
  model and unstructured equity covariance matrices.
\newblock {\em Journal of Risk}, 10:3--30.

\bibitem[Fama and French, 1993]{Fama-93}
Fama, E. and French, K. (1993).
\newblock Common risk factors in the returns of stocks and bonds.
\newblock {\em Journal of Financial Economics}, 33:3--56.

\bibitem[Ledoit and Wolf, 2004]{Ledoit-Wolf.2003}
Ledoit, O. and Wolf, M. (2004).
\newblock Honey, i shrunk the sample covariance matrix.
\newblock {\em Journal of Portfolio Management}, 4(30):110--119.

\bibitem[Marcenko and L.A.Pastur, 1967]{MarcenkoPastur}
Marcenko, V. and L.A.Pastur (1967).
\newblock Distribution of eigenvalues for some sets of random matrices.
\newblock {\em Math. USSR-Sb}, 1:457--483.

\bibitem[Potters et~al., 2005]{PotterBouchaudLaloux}
Potters, M., Bouchaud, J.-P., and Laloux, L. (2005).
\newblock Financial applications of random matrix theory: Old laces and new
  pieces.
\newblock {\em Acta Phys. Pol.}, B36:2767.

\bibitem[Tsay, 2005]{Tsay-05}
Tsay, R. (2005).
\newblock {\em Analysis {o}f Financial Time Series}.
\newblock 2nd edition.

\bibitem[Wishart, 1928]{Wishart}
Wishart, J. (1928).
\newblock The generalized product moment distribution in samples from a normal
  multivariate population.
\newblock {\em Biometrika A.}, 20:32–43.

\bibitem[Zumbach, 2004]{Zumbach.LongMemory}
Zumbach, G. (2004).
\newblock Volatility processes and volatility forecast with long memory.
\newblock {\em Quantitative Finance}, 4:70--86.

\bibitem[Zumbach, 2006]{Zumbach.RM2006_fullReport}
Zumbach, G. (2006).
\newblock The riskmetrics 2006 methodology.
\newblock Technical report, RiskMetrics Group.
\newblock available at: http://www.riskmetrics.com/publications/techdoc.html.

\bibitem[Zumbach, 2008]{Zumbach.inferenceOnProcesses}
Zumbach, G. (2008).
\newblock Inference on multivariate {ARCH} processes with large sizes.
\newblock Technical report, RiskMetrics Group.
\newblock Forthcoming.

\end{thebibliography}

\end{document}